# Reproducibility and scientific interpretation in the age of AI: consilience in biological systematics, ecology, and molecular biology

CHARLES MORPHY D. SANTOS[1], LUCIANA CAMPOS PAULINO[2], MICHAELLA P. ANDRADE[3], GABRIEL TOGNELLA-POCCIA[4] & JOÃO PAULO GOIS[5]

1- *Centro de Ciências Naturais e Humanas, Laboratório de Sistemática e Diversidade, Universidade Federal do ABC, Av. dos Estados, 5001. Bairro Bangu, CEP 09210-580, Santo André/SP, Brazil. E-mail: charlesmorphy@gmail.com. ORCID: https://orcid.org/0000-0001-5577-0799.*

2- *Centro de Ciências Naturais e Humanas, Universidade Federal do ABC, Av. dos Estados, 5001. Bairro Bangu, CEP 09210-580, Santo André/SP, Brazil. E-mail: luciana.paulino@ufabc.edu.br. ORCID: https://orcid.org/0000-0003-2421-5236*

3- *Centro de Ciências Naturais e Humanas, Laboratório de Sistemática e Diversidade, Universidade Federal do ABC, Av. dos Estados, 5001. Bairro Bangu, CEP 09210-580, Santo André/SP, Brazil. E-mail: michapereirandrade@gmail.com. ORCID: https://orcid.org/0000-0001-8717-7746*

4- *Centro de Ciências Naturais e Humanas, Universidade Federal do ABC, Av. dos Estados, 5001. Bairro Bangu, CEP 09210-580, Santo André/SP, Brazil. E-mail: gabrielpoccia@gmail.com. ORCID:https://orcid.org/0000-0002-4033-7631*

5- *Centro de Matemática, Computação e Cognição, Universidade Federal do ABC, Av. dos Estados, 5001. Bairro Bangu, CEP 09210-580, Santo André/SP, Brazil. E-mail: jpgois@gmail.com. ORCID: https://orcid.org/0000-0002-9437-6943*

**Abstract**

Achieving complete reproducibility in science, particularly in research fields such as biodiversity, is challenging due to analytical choices, bias and interpretation. Here, we examine examples of reproducibility in biological systematics, ecology, and molecular biology. To mitigate the impact of interpretation and analytical choices, Artificial Intelligence (AI) has provided potential tools. In the present work, while emphasizing the need for methodological rigor and transparency, we acknowledge the role of interpretation in activities such as coding biological characters and detecting morphological patterns in nature. We explore the opportunities and limitations associated with the synergy between big data and AI in molecular biology, emphasizing the need for a more comprehensive and integrated approach based on dataset quality and usefulness. We conclude by advocating for AI-based tools to assist biologists, reinforcing consilience as a criterion for scientific validity without hindering scientific progress.

## 1. Introduction

Discovery in science includes formulating ideas, conducting controlled experiments, identifying new entities through observation, recognizing properties of well-established entities, and developing new principles, laws, and theories (Arabatzis 2006). Scientific discovery relies on testing hypotheses with empirical evidence. A strong empirical foundation allows theories to be refined and validated by the scientific community. Here lies a fundamental issue in science: the ability to independently reproduce results using the same empirical evidence and analogous experimental designs.

Plesser (2018) highlights ongoing confusion and debate over the terms repeatability, replicability, and reproducibility across scientific disciplines. According to the Association for Computing Machinery (2020), repeatability is related to situations where the measurements can be obtained with precision by the same team using the same measurement procedures and measuring systems under the same operating conditions, in the same location in multiple trials; replicability is similar, but considers different teams, with the same experimental setup of the original proposal; and reproducibility relates to situations where independent groups can obtain the same results as the original but with different measuring system, in a different location on multiple trials. The term reproducibility captures the challenge of balancing interpretation with replicability.

The quest for reproducibility faces significant obstacles in biological systematics, ecology, and molecular biology, which rely heavily on observation and decisions regarding data processing, model selection, and interpretation. Recently, Gould (2025) presented a large-scale study on scientific reproducibility, which examined how over 200 biologists interpreted the same ecological datasets yet arrived at widely different conclusions. The primary cause of the discrepancies was not the data but the diversity of the researcher's analytical choices. This finding underscores a critical issue in science: even when working with identical datasets, researchers bring distinct theoretical frameworks, methodological preferences, and personal biases to their analyses.

Reproducibility concerns have been increasing, particularly with the growing use of Artificial Intelligence (AI)-based algorithms in biological research (Castelvecchi 2016, Hutson 2018, Ball 2023, Mersha et al. 2024, Frazier and Song 2025). The so-called opacity of AI models complicates the reconstruction of the processes that lead to specific research results. A possible solution is Explainable AI (XAI), an approach that provides tools to identify the decision-making processes of AI models, thereby improving their interpretability and reproducibility (Minh et al. 2022, Longo et al. 2024).

Despite technological data collection and analysis advancements, reproducibility remains a key concern in biological systematics, ecology, and molecular biology. Jenkins et al. (2023) emphasized that making raw data, standardized and detailed metadata, and analysis scripts available is essential for study validation, data synthesis, and scientific progress in ecological and evolutionary research.

One of the remaining gaps in biology research is understanding how analytical choices impact scientific conclusions, even when researchers use identical datasets. A second gap is the trade-off between dataset quantity and quality in an era dominated by big data and AI applications. Large, uncurated datasets frequently lack essential metadata, limiting their usefulness and reproducibility in biodiversity studies and molecular research.

Total objectivity is not reachable in scientific inquiry. Nevertheless, difficulties in achieving reproducibility should not hinder scientific progress. Instead, we should acknowledge the inherent subjectivity and search for strategic tools to maximize transparency and open science practices. In AI-based biological studies, researchers' attention must be doubled so that the results correspond to empirical evidence, be it species for taxonomy, morphological and molecular characters for phylogenetic analyses, or behavioral attributes for ecology studies. Ensuring AI reproducibility requires transparent documentation of datasets, algorithms, and preprocessing steps.

AI can standardize character coding and automate pattern recognition, enhancing consistency. However, its opacity and occasional reliance on poorly annotated datasets pose challenges to reproducibility. This paper examines how analytical choices and interpretation influence scientific reproducibility across biological disciplines and explores the potential of AI-driven methodologies to enhance reproducibility while mitigating inherent biases. We analyze case studies in biology research, advocating for a rigorous framework in data collection, annotation, and methodological transparency.

We also propose a path forward for integrating AI into the biological sciences by drawing on the consilience of inductions, i.e., the convergence of diverse scientific findings toward a unified explanation (cf. Whewell 1847). This framework emphasizes how AI models, although often developed with narrowly defined objectives, frequently yield insights far beyond their initial scope. Such unexpected explanatory power indicates a form of epistemic robustness, reinforcing the validity of the underlying theories and hypotheses. We suggest that AI can transform biological understanding through integrative, cross-disciplinary discovery by aligning AI's predictive and explanatory capabilities with this philosophical principle.

## 2. Reproducibility in biological systematics

Phylogenetic trees to show species' evolutionary relationships require significant interpretation. This is a concern since most morphology-based analysis is still based on human observation before the matrix coding. A mandatory step in morphological phylogenetic analysis is the transformation of raw morphological observations into lists of characters and character states, which serve as hypotheses about the shared evolutionary origin of these traits. These character-state hypotheses are then organized into character vs. taxon matrices, which are inputs to software such as TNT (Goloboff and Catalano 2016) and Mesquite (Maddison and Maddison 2023). The results are phylogenetic trees—dendrograms showing the evolutionary relationships among the biological groups under investigation. None of the steps above are AI-driven in these traditional methods currently available.

Character coding translates morphological diversity into comparable variables but remains subjective. Systematists may apply different schemes, producing varied matrices and phylogenetic results. Analytical criteria, such as weighting certain characters more than others, further shape these outcomes. Consequently, systematists may formulate distinct phylogenetic hypotheses even when working with identical biological specimens.

Examine the following case considering five hypothetical species of the genus *Chrysopilus* Macquart (Diptera, Rhagionidae). All of them present variation in their gonostylus, a male genital structure used in grasping female abdomens during reproduction: *C.* sp. 1 (Fig. 1A) has only a tiny bud articulated with the gonocoxite, a genital structure derived from appendages of segment IX; *C.* sp. 2 (Fig. 1B) has a rounded and setose gonostylus; *C.* sp. 3 has a rounded and bare gonostylus; *C.* sp. 4 has a subsquare and setose gonostylus; and *C.* sp. 5 has a subsquare and bare gonostylus. There are many ways to code characters based on observation of the gonostylus, as presented in Fig. 1:

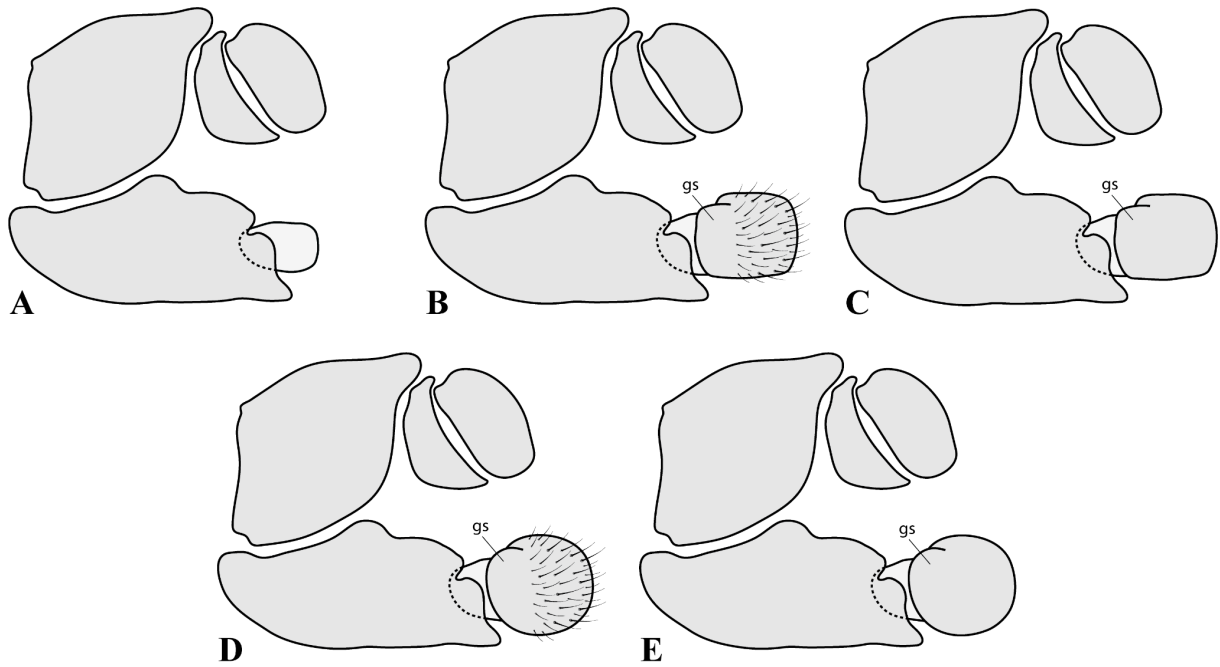

**Figure 1.** Different expressions of the gonostylus (gs) in hypothetical species of *Chrysopilus* (Diptera, Insecta). **A.** Absent. **B.** Rounded and setose. **C.** Rounded and bare. **D.** Subsquare and setose. **E.** Subsquare and bare.

**Character coding A**

1. Gonostylus: (0) absent; (1) rounded and setose; (2) rounded and bare; (3) subsquare and setose; (4) subsquare and bare

**Character coding B**

1. Gonostylus: (0) absent; (1) rounded; (2) subsquare
2. Gonostylus: (0) absent; (1) setose; (2) bare

**Character coding C**

1. Gonostylus: (0) absent; (1) present
2. Form of the gonostylus: (0) rounded; (1) square
3. Setae in the gonostylus: (0) absent; (1) present

**Character coding D**

1. Gonostylus: (0) absent; (1) present
2. Rounded shape of the gonostylus: (0) absent; (1) present
3. Squared shape of the gonostylus: (0) absent; (1) present

4. Setae in the gonostylus: (0) absent; (1) present

Each coding approach treats the observable features differently. When the resulting matrices are analyzed through parsimony, the results also differ (Fig. 2). In summary, even the coding of a simple morphological attribute may impact the final phylogenetic pattern. Interpretation and analytical decisions also affect the output of molecular-based phylogenies, and numerous forms of standardization have been discussed (Shakya et al. 2020).

Indeed, we should expect that independent researchers with the same raw evidence, i.e., biological species, analyzed with the same criteria and software used in the original research, would obtain the same results—after all, reproducibility does exist in phylogenetic systematics—but different interpretations of the analytical unities, such as the observed specimens, are inherent to systematics.

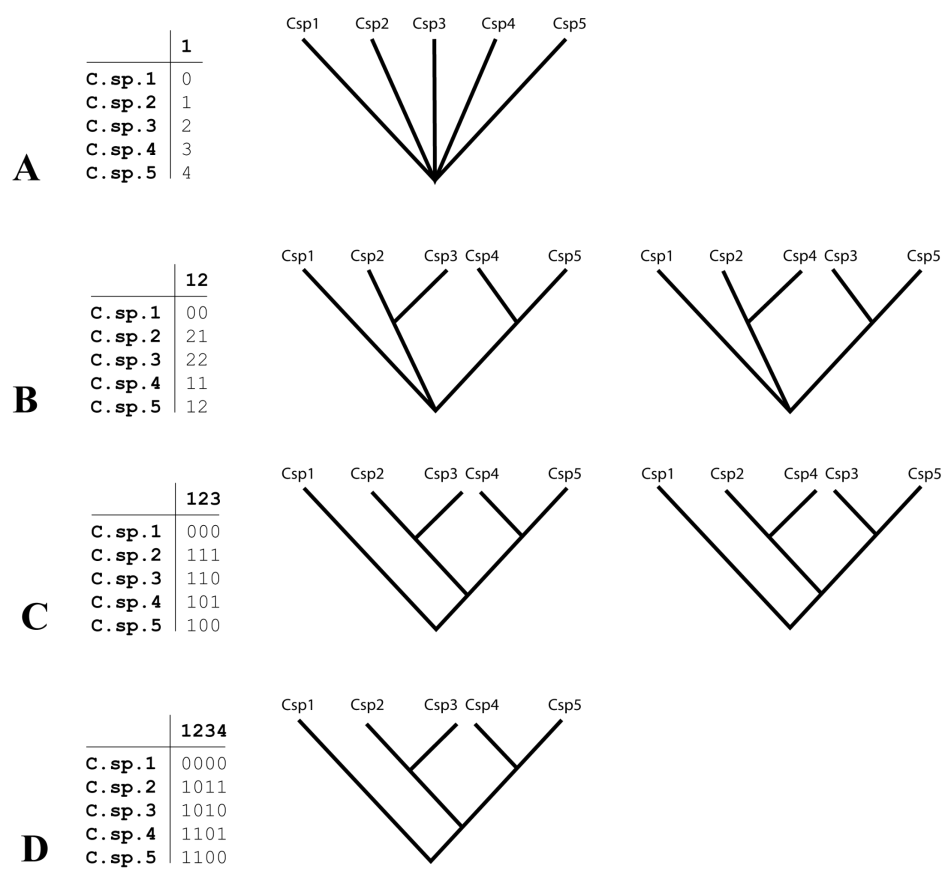

**Figure 2.** Different coding schemes **(A–D)** based on varying interpretations of the same attribute, the gonostylus in hypothetical *Chrysopilus* species, lead to distinct matrices and phylogenetic hypotheses.

One practical approach to address reproducibility in taxonomy and phylogenetic analysis is expanding the primary dataset used in morphological studies by including more biological material during character observation and coding. Furthermore, by incorporating data from microstructural observations, such as those obtained through scanning electron microscopy, the resolution and detail of morphological characters are significantly enhanced, which may result in more precise character definitions. Additionally, training and capacity-building initiatives for biological systematists specialized in interpreting and describing morphological traits are necessary since traditional taxonomy expertise has declined (Santos and Carbayo 2021).

AI-assisted tools can enhance reproducibility in taxonomy by automating character observation and classification. These tools can standardize taxonomic identification if trained on diverse morphological datasets—including microscope photographs, illustrations, and electron microscopy images—though validation across datasets remains an issue. To tackle the problem of taxonomic identification of biological species, Valan et al. (2019), for example, employed transfer learning from convolutional neural networks (CNNs) through an off-the-shelf model featuring more than one hundred million parameters, pre-trained on a dataset of over one million annotated images. To be applied in insect species identification, Valan and collaborators performed the feature transfer, an approach of transfer learning suitable to smaller datasets comprising a few hundred to around three thousand images. By leveraging predefined rules and large comparative datasets, AI can assist in minimizing human bias and improve consistency in character coding.

Furthermore, using AI-based automated tools to propose heuristic hypotheses of homology before the coding process is also an alternative to reduce bias and the effect of interpretation in character coding. In this context, imageomics is an emerging field that applies AI tools grounded in biological knowledge to analyze image data in order to identify patterns and extract insights about traits and relationships at the individual, population, and species levels (Kline et al. 2023). Pollock et al. (2025) suggest that future AI applications in biological systematics may allow the inference of trait dendrograms directly from organism images, including three-dimensional representations of fossils, offering new ways to generate testable hypotheses about phylogenetic relationships. The authors also highlight the potential of AI to analyze high-resolution 3D scans of natural history specimens to investigate trait evolution and reconstruct ancestral states. In their study on birdwing butterflies, Cuthill et al. (2024) show that AI can detect subtle phenotypic differences in wing shape and color—traits often difficult for humans to distinguish. Their findings reveal complex phenotypic variation patterns, including well-documented male traits linked to sexual selection and less-explored variation in females. These results underscore the potential of AI to detect biologically meaningful phenotypic variation that extends beyond the patterns typically recognized by taxonomists and systematists.

Obviously, we do not advocate that the simple creation of unsupervised models based on unlabelled taxonomic data will solve any reproducibility concern in biological systematics. AI will not substitute taxonomists or phylogeneticists but may assist biologists as a computational co-pilot in standardizing character definitions and facilitating data integration. In this case, biologists and AI are specialists working toward a holistic framework to pursue consilience, as discussed below.

### 3. Reproducibility in ecology

Poor reproducibility in ecological tests weakens results, particularly when uncontrolled environmental conditions affect experimental design (Andrade et al. 2023), making hypothesis testing more challenging.

As in biological systematics, data interpretation and biased analytical choices in ecology hinders even collaborative work. Divergent conclusions within the same team and experimental setup often stem from personal interpretations of observed data, as all observation is inherently theory-laden. Hence, the conclusion of Gould et al. (2023) is valid: when there is a diverse set of plausible analytical options, no single analysis is the complete answer to a research question. In this sense, reproducibility is an elusive target.

In ecology, results are expected to be at least partially reproducible by other researchers in different contexts. In octopus research, general body patterns recur across different contexts. However, these patterns are influenced by many hard-to-track variables, making comparative analyses difficult. Researchers using the same analytical methods may still obtain varying results (e.g., How et al. 2017). Key variables affecting ecological research include environmental fluctuations, seasonality, biotic interactions, and methodological choices.

The following example shows how individual biases impact behavior studies in *Octopus insularis* (Figure 3). The variety and intensity of octopuses' body patterns, defined after photograph coding of skin components, can lead to discrepancies among researchers analyzing the same specimens. Consider the "blotch" pattern, which is defined by the presence of circular white spots and dots usually on the octopus' mantle and arms (Andrade et al. 2023). Here, variants 3A, 3B, 3C, and 3D of the "blotch" pattern are notable, with 3B and 3D showing dark bars on the arms of the octopuses, unlike 3A and 3C. The "blotch" pattern classification can thus vary; it can be limited to instances without bars on the arms or be defined by the presence of specific chromatic components on the mantle.

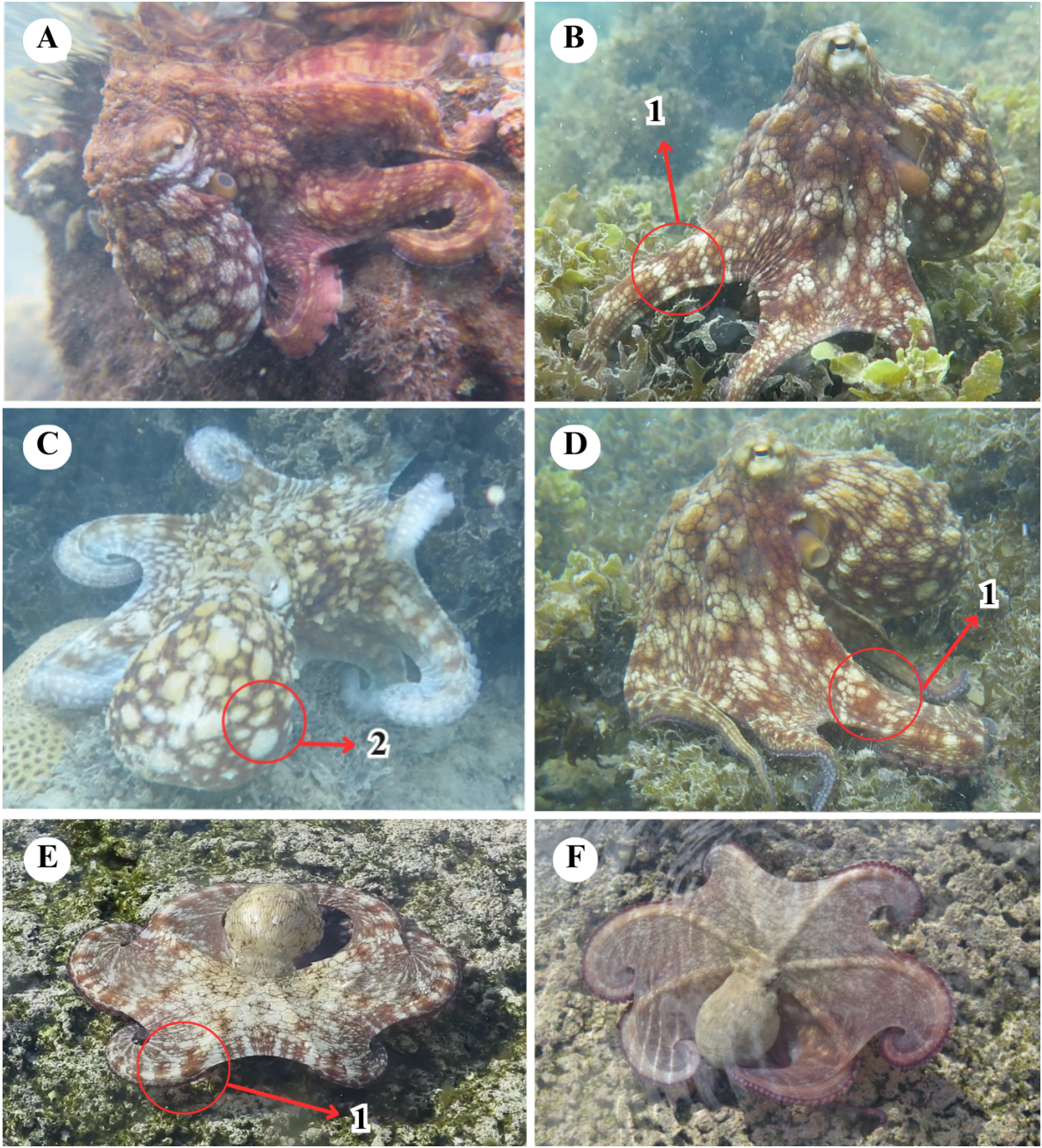

**Figure 3.** Chromatic components in wild *Octopus insularis*. **A–D.** Variations of the blotch pattern. **E.** Mottle pattern. **F.** Deimatic pattern. **1.** Darker brown bars in arms. **2.** Blotches (non-uniform pale circular area) and spots (uniform pale circular area in 360 degrees of its dimensions) in the mantle. Photos by Michaella P. Andrade.

Figure 3E shows the mottle body pattern, when the octopus has a skin pattern varying from reddish to brown. It has a white "V" just below the eyes and bars alternating between light and dark tones on the arms. In Figure 3F, the deimatic display shows the octopus with mostly pale skin, arched arms, and open webs. Although the body posture displayed by the octopuses in 3E and 3F is similar, only 3E has contrasting bars on its arms, further complicating the classification, as some researchers may interpret this as a deimatic pattern. Such subjective interpretations significantly impact the analysis of octopuses' behavior, as illustrated by a video analysis lasting 2 minutes and 1 second that examines the duration of body patterns, with coding schemes differing based on the presence or absence of dark arm bars (Figure 4).

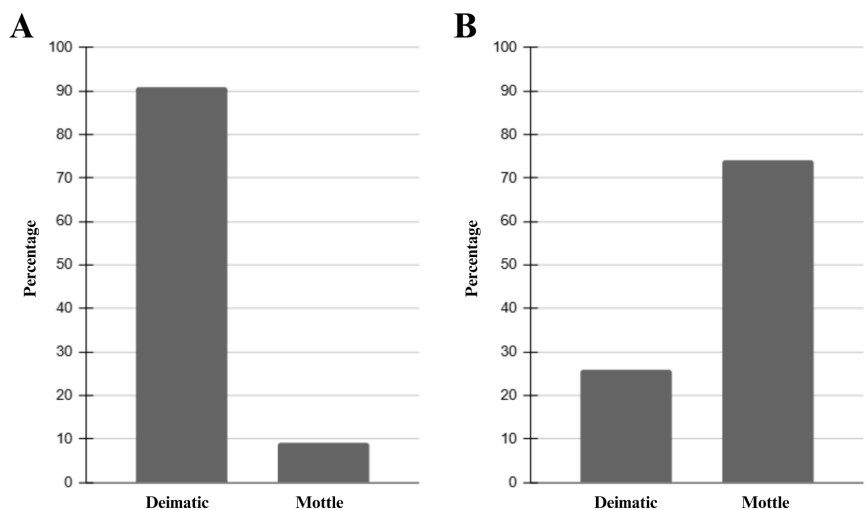

**Figure 4.** Percentage of body patterns in a video 2 minutes and 1-second long. **A.** Coding that considers the bars as markers of the mottle pattern. **B.** Coding that prioritizes the postural pattern with arched arms and open webs as a marker of deimatic pattern, regardless of the presence of the dark bars on the arms.

The subjectivity involved in interpreting octopus displays poses challenges for ecological and behavioral studies, particularly those examining the frequency, duration, and latency of these visual cues across various contexts and environmental conditions (Andrade et al. 2023). Interpretative discrepancies are especially pronounced during transitional moments between distinct body patterns. Training researchers to classify these visual signals based on detailed descriptions can help reduce individual biases, although some analytical variability will inevitably persist. Such different interpretations arise from researchers' diverse scientific perspectives and intellectual backgrounds, limiting the reproducibility and consistency of ecological findings.

With the popularization of AI-based pattern recognition and object identification tools that allow confident identification, the likelihood of interpretative divergence among researchers is expected to decrease. Tools such as YOLO-SAG (Chen et al. 2024), an improved AI model based on YOLOv8n for wildlife target detection that balances speed and object detection accuracy, are a pathway towards automatizing biodiversity monitoring systems. DeepLabCut (Mathis et al. 2018) and DeepEthogram (Bohnslav et al. 2021) are other AI-based, high-performance automated animal tracking and sorting methods. Here, *in situ* imageomics (Kline et al. 2023) can also be a powerful set of AI-based tools to collect images directly in the field and integrate them with knowledge bases, allowing for the characterization of wildlife populations.

The use of image data to answer ecological questions is commonplace within ecology research (Trotter et al. 2025). Still, there is room for advancing AI models to automate processes that have been manually executed for decades partially. Pollock et al. (2025) emphasize that AI is expected to increasingly uncover traits beyond the limits of human perception, advancing its capabilities by integrating diverse data modalities—such as two-dimensional and three-dimensional imaging and dynamic video formats—and revealing connections among them, thus enabling the detailed capture of organism behavior in natural environments in a truly consilient process such as discussed below. The conceivable revolution of animal behavior research in the next few years lies in the ability of AI to extract meaningful insights from massive datasets of photographs and video recordings (Saoud et al. 2024).

As Pollock et al. (2025) discuss, AI is increasingly being used to enhance traditional models of species interactions, such as plant-pollinator, predator-prey, and host-parasite networks, by integrating data sources like trait matching, environmental DNA, species co-occurrence, and network structures. Even in the absence of detailed trait data, AI can extrapolate interaction patterns using graph dimensionality reduction techniques, which create compact representations of interaction networks (or metawebs) and can be transferred across taxa. Strydom et al. (2023) suggest that maximizing available information through graph embeddings is an effective way for improving metaweb predictions. These machine-learning techniques map discrete graph structures into a continuous space (Chami et al. 2022), offering a more efficient alternative to exhaustive lists of species interactions, which is currently the standard practice in ecology. Thus, AI is expected to advance the analysis of ecological interaction networks and even predict extinction cascades by identifying species losses likely to affect network structure and function disproportionately (Pollock et al. 2025).

We reinforce Santos and Gois (2025)'s opinion: in biodiversity studies, AI developments should converge with biologists' demands. Concurrently, biologists should standardize species description templates, behavior categories and definitions of ecological interactions, and metadata that allow the effective training of AI-based tools. This synergy aligns with the active search for consilience advocated here. It will improve the accuracy of automated analyses and facilitate the integration of AI into ecological research.

### 4. Reproducibility in molecular biology

In molecular biology, as in other science fields, we live in the era of big data (Pal et al. 2020, Cremin et al. 2022). The emphasis on large datasets has led to the widespread conception that genomes represent the ultimate achievement of biology. In fact, the possibility of converting vast volumes of data into biological predictive knowledge has been questioned since the 2000s (Lewontin 2001). Unfortunately, the ease of obtaining data has contributed to perpetuating a cycle in which the solution is sought by generating more and more data. We emphasize the need for a more comprehensive and integrated data collection approach towards reproducibility.

Automated DNA sequencers and sequencing platforms made it feasible to determine DNA sequences of entire genomes, establishing the field of Genomics and being the base of the so-called Postgenomics (Hocquette 2005). The achievements of Genomics contributed to the perception that DNA sequences would be the starting point for any biological investigation. In the early 2000s, the GenBank and WGS NCBI databases contained approximately 5.8 billion nucleotides; over 22 years, this number has exceeded 17 trillion, according to the National Center for Biotechnology Information (2025). Despite vast genomic datasets, a fragmented approach has often failed to provide meaningful biological insights.

Large, unannotated databases often lack metadata, limiting their scientific utility. Genomic datasets that include detailed descriptions—such as the host organism, the environmental source of the specimen, spatial-temporal data, tissue collection methods, DNA extraction techniques, and sequence library preparation—can yield far more substantial scientific insights (Quiñones et al. 2020). Without adequate metadata, datasets lose immediate value and become difficult to reuse in future research. Nonetheless, inadequately annotated data impedes the confidence training of AI models, thereby complicating the development of models able to discern patterns within genomic databases.

In the case of sequencing data, the value lies in how data can provide insights into scientific questions and substantiate testable hypotheses. Consequently, large datasets must exhibit precision and objectivity to ensure reproducibility. Errors stemming from inaccurate annotations in databases, data preprocessing mishandling, or poor communication of methods can invalidate scientific ideas and propositions. The following example illustrates this issue. Poore et al. (2020) analyzed microbial genomes and transcriptomes from a large cohort of cancer patients. It proposed a potential cancer diagnostic tool based on microbial signatures. However, the conclusions were questioned by Gihawi et al. (2023). The rebuttal paper argued that a substantial amount of human DNA was incorrectly labeled as microbial sequences in the databases, biasing the machine-learning model and resulting in inaccurate signatures. The reevaluation of raw and normalized taxonomic assignment data unveiled the actual count of bacterial sequences, which was significantly lower than the initial analysis suggested. These flaws affected numerous subsequent studies that relied on the data from the original paper.

Lack of standardization and insufficient or incorrect information regarding gene and protein functions, sample origin, and the relationships between genotypes and phenotypes worsen as data volume increases (Philippi and Köhler 2006, Schnoes et al. 2009). Researchers have recognized the need for structured metadata to address gaps in open omics datasets for decades (Quiñones et al. 2020, Rajesh et al. 2021, Sasse et al. 2022). Institutional avenues could play a significant role in this process. Today, numerous scientific journals stipulate that the authors deposit their data in public repositories, which is a responsible practice. Journals have contributed to the community's adherence to ethical practices since they started demanding approval from Institutional Review Boards for publishing papers on human subjects and animal models. Besides, it would be important to interconnect databases, establishing interfaces between primary, secondary, and specialized databases. Ensuring that biological databases align with the FAIR (Findable, Accessible, Interoperable, Reusable) principles (Wilkinson et al. 2016, Jenkins et al. 2023) is also an essential step toward optimizing the use of large datasets in the life sciences.

As Baykal et al. (2024) discussed, rapid technological advancements may introduce biases, inconsistencies, and variability that complicate genomic-related research. The authors emphasize that, given its critical role in areas such as medicine, genomic reproducibility must be prioritized through rigorous evaluation of bioinformatics tools, adherence to best data collection practices, and standardized guidelines to ensure the accurate translation of genomic findings into clinical applications. Training supervised models rely on datasets with annotated labels, which supply the necessary information to guide the learning process and estimate a function or conditional distribution over target variables from given inputs (Wang et al. 2023). AI-based methods in molecular biology can only benefit from these careful protocols: a milestone in the understanding of the human genome was the development of deep learning models that can predict phenotypes from genotypes, which advanced our ability to understand and design genomics sequences (Dalla-Torre et al. 2025). In this sense, although subjectivity should be taken for granted in any science, careful curation of "omics" datasets is needed to improve reproducibility, particularly in light of recent developments in AI-based algorithms for learning from molecular data.

Despite these challenges, scientific inquiry does not depend entirely on perfect reproducibility. Instead, consilience—the convergence of independent evidence—offers a complementary approach to validating scientific findings. The following section explores how AI and other tools contribute to this process.

## 5. Towards consilience

While the truth remains unattainable, the convergence of multiple scientific outcomes supports the idea that the hypothesis under scrutiny reveals something about the underlying reality. This illustrates the concept of consilience, which is particularly useful in biological sciences (Capellari and Santos 2012, Santos and Capellari 2009, White 2024).

Consilience, initially introduced by Whewell (1847), refers to situations in which a causal explanatory theory successfully explains phenomena not considered initially when the theory was formulated. According to Capellari and Santos (2012), consilience can serve as a criterion to assess the validity of a scientific theory. Even though science may not achieve "truth", the likelihood of unrelated phenomena being integrated into the same scientific theory purely by chance is low. Such convergence provides robust support for scientific claims, even without perfect reproducibility.

This does not mean that we should abandon reproducibility. Partially repeatable, replicable, and reproducible studies, as long as they converge to similar answers, are acceptable as steps in the progress of science. In a phylogenetic analysis, for example, different interpretations of the same empirical evidence—in this case, the observed specimens—even though leading to divergent hypotheses of evolutionary relationships, may share consistent monophyletic groups. These phylogenetic patterns are the foundation for more detailed studies using different sets of evidence, which can be subject to various interpretations. This continuous process allows us to refine hypotheses, bringing us closer to the actual evolutionary history of the biological group under investigation.

The example from ecological studies we are addressing here is the behavior of octopuses in natural, uncontrolled environments. Even if two researchers from the same team interpret octopuses' body patterns differently, we should not discard such a partially repeatable study. Furthermore, *in natura* conditions are difficult to replicate, making it unlikely that other teams will achieve similar results. Still, even a single observation in the field can illuminate the comprehension of the evolution of a given animal behavior in highly variable environments. In this case, careful documentation, availability of data in open and accessible repositories, and accurate communication of results are *sine qua non* conditions for ensuring that hypotheses are more reliable.

It is worth mentioning that AI integrates insights from largely unrelated data to create predictive models that approximate reality, though without fully capturing it. AI-based tools may reveal patterns within the data from multiple sources not explicitly anticipated during the initial hypotheses formulation, transcending its original scope. As stated by White (2024), consilience is an integrative approach to scientific problems that cut across disciplines and cultures. In a certain sense, AI predictive models align with Whewell's (1847) notion of consilience. It seems no exaggeration to say that AI-driven science mirrors the scientific process itself: continuously refining models that, while never fully capturing the ultimate truth, progressively align with the underlying structure of nature (Santos and Capellari 2009).

Ruse (1979) states that a reliable scientific theory should be consilient. AI-based models are lauded for their ability to identify patterns, infer relationships, and make predictions across varied domains. In philosophical terms, this capacity resonates with Whewell's concept of the consilience of inductions. Whewell's optimistic notion provides a framework for understanding the scientific reliability of AI models: they show consilience by generating explanatory power across domains far removed from their original design.

According to Whewell (1847, p. 77-78): "the evidence in favor of our induction [our theory or hypothesis] is of a much higher and more forcible character when it enables us to explain and determine cases of a kind different from those which were contemplated in the formation of our hypothesis. The instances in which this has occurred, indeed, impress us with a conviction that the truth of our hypothesis is certain. No accident could give rise to such extraordinary coincidence." Hence, for Whewell, a scientific hypothesis (H) gains exceptional epistemic value when it successfully explains not only the phenomena it was originally intended to address (G1) but also other, previously unrelated or unexpected phenomena (G2, G3, Gn) (Figure 5). The more a hypothesis extends its explanatory power beyond its initial scope, the more it should be trusted—not simply as a plausible account but potentially as one that has grasped a real underlying causal mechanism, or what Whewell called a *vera causa*. An attractor, which in dynamical systems represents a set of states towards which a system tends to evolve, never intersecting themselves but confined to a finite space (Gleick, 2008), can be used as a visual metaphor for scientific research: hypotheses may gravitate towards the ultimate truth about an issue. Even though the truth remains unattainable, the convergence of multiple research outcomes, not identical but similar, supports the idea that the hypothesis under scrutiny reveals something about reality.

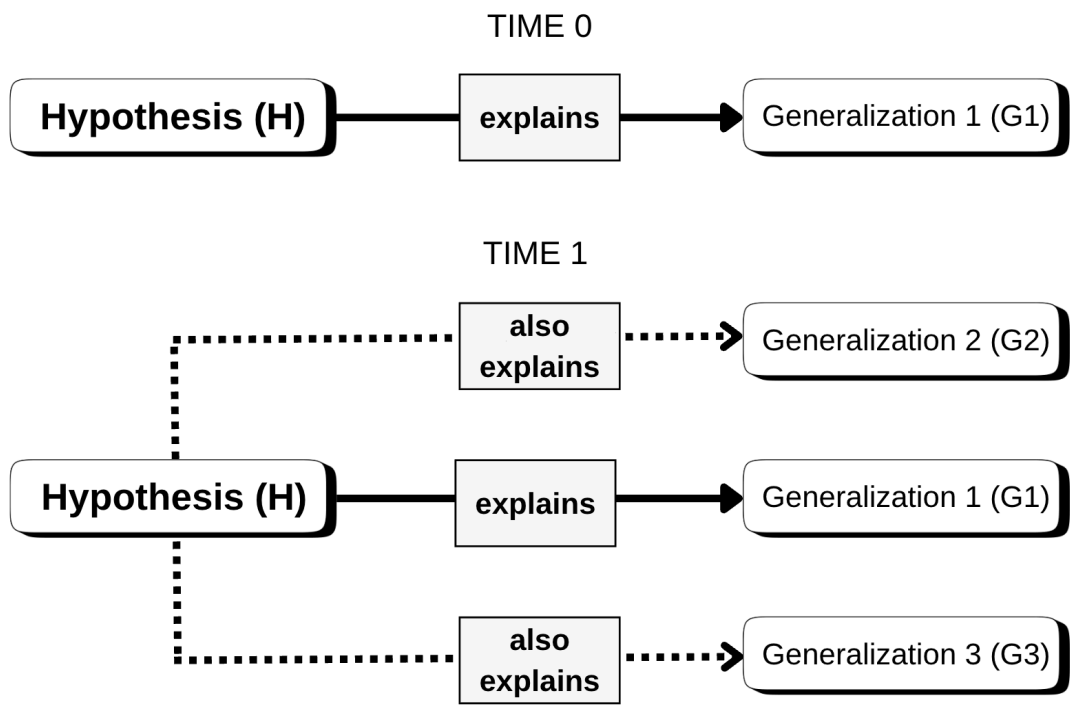

**Figure 5.** Schematic representation of consilience. At time 0, the Hypothesis (H) explains only Generalization 1 (G1). At time 1, G2 and G3, which are not expected to be explained by H, are actually explained by the Hypothesis. H can be reinterpreted and stretched to account Gn, or Gn reinterpreted under H. Hypothesis H is consilient, with increased support (modified from Santos and Capellari 2009).

Rather than evaluating a hypothesis solely by its ability to fit predefined observations, consilience values a theory's unanticipated explanatory reach. Such cases acquire a higher degree of epistemic support since they would reflect an unlikely coincidence if the theory were false (Santos and Capellari 2009, Capellari and Santos 2012). If a single model can account for disparate phenomena, the simplest explanation may be that the model is capturing something fundamental about the world.

AI-based models start by training on large datasets to predict specific outcomes, such as identifying color patterns in photographs or animal behaviors caught on camera. Initially, these models are designed for a narrow task (G1). However, once trained, they may be repurposed and extended to new, often unforeseen domains (G2, G3, Gn), sometimes with minimal adjustments. This unexpected generalization mirrors the structure of Whewellian consilience.

For instance, a natural language processing model like GPT was initially trained to predict the next word in a sentence based on a vast corpus of raw text (Radford et al. 2008). However, it is effective at a broad spectrum of tasks: translation, summarization, dialogue generation, coding assistance, preparing taxonomic descriptions based on photos, scientific paper reviewing, and clinical, educational, and research work in medicine (Santos and Gois 2023, Thirunavukarasu et al. 2023, Yenduri, 2024). These capabilities were not explicitly programmed or anticipated in the training objectives. In Whewellian terms, GPT-style models were designed to explain or predict G1 (word sequences) but have shown robust explanatory capacity over G2, G3, and beyond. The same is valid for computer vision models trained to identify cats and dogs in images (Tripathy and Singh 2022) and later deployed in medical imaging to detect tumors or satellite imagery to monitor deforestation (Poornam and Angelina 2024, Careli et al. 2024). Again, their initial explanatory task (G1: object classification) expands into qualitatively different domains (G2, G3, Gn) that were not considered initially. This extension is a sign of underlying pattern recognition that crosses epistemic boundaries. Consider AlphaFold, an AI developed to predict the three-dimensional structure that a protein will adopt based on its amino acid sequences (Jumper et al. 2021). Initially, this model was trained on known protein-folding data (G1). Still, its predictions have since opened new pathways in drug discoveries, vaccines, enzymatic processes, and determining the rate and effect of different biological processes (G2, G3, Gn), allowing the identification of unknown structures and suggesting novel biological functions (Desai et al. 2024). These expansions were not explicitly coded in the initial goals of AlphaFold but emerged from the model's internalization of general principles embedded in the data.

The implications of AI's consilient behavior resonate with Whewell's epistemology. Whewell saw scientific discovery as a synthesis of "facts" and "ideas," where conceptual innovation leads to unifying disparate observations under a common explanatory umbrella.

AI performs an analogous function: mapping data into high-dimensional spaces captures latent structures that unify seemingly unconnected events.

As mentioned before, a key issue of Whewell's concept is its association with "truth." However, as Santos & Capellari (2009) and Capellari & Santos (2011) elaborate, despite not guaranteeing truth, consilience may be seen as a "criterion of reality." A theory or model that unifies diverse domains is likely tracking something real, even imperfectly. Translating this to AI, we may not say that a model like GPT or AlphaFold "discovers truths" in the classical sense. Nevertheless, their consilient success indicates a reliable representation of complex structures in the world. This provides a compelling epistemic justification for trusting the predictions of these models, even when their internal operations are opaque.

Moreover, AI models often outperform hand-crafted scientific models since humans struggle to discern between high-dimensional and nonlinear relationships from which AI can draw insights (e.g., Cuthill et al. 2024). AI models' success at generalizing to unforeseen phenomena invites the development of new theoretical insights in a feedback loop between consilient output and human understanding.

In the not-distant past, theory-building was primarily a human cognitive activity guided by creativity, intuition, and empirical grounding. The rise of computational science challenges this perspective, as its methods increasingly displace humans from the center of the epistemological process (Humphreys 2009). Now, AI-based models producing outputs that mirror the behavior of consilient theories raise questions about the nature of scientific explanation itself. AI models could be proto-theories, as empirical engines that generate generalizations that later gain theoretical interpretation or additional evidence, turning the new theory or hypothesis into a merge of irreproducible insights, partially repeatable and reproducible experimental designs. It is an open question that future epistemic developments of AI-based science could embed.

The Whewellian perspective also invites a reevaluation of theory interpretability. While reproducibility remains a critical concern, as discussed here, consilience offers a complementary epistemic measure, in the sense elaborated by Santos and Cappelari (2009) in the biogeographical context. Even if a model's internal workings are not discernable, its capacity to explain multiple kinds of phenomena increases its trustworthiness. Regarding biodiversity and other biological sciences, as Pollock et al. (2025) discuss, the power of AI to synthesize scientific knowledge from various sources, coupled with the increasing availability of sequenced genomes, could lead to the development of new evolutionary foundation models.

Finally, the concept of Computational Reliabilism (CR) (Durán and Formanek 2018, Russo et al. 2024) offers a helpful framework for evaluating the reliability of computational processes. A key indicator of CR is robustness analysis, which suggests that if a sufficiently diverse set of models consistently produces the same result, the real-world phenomenon also likely exhibits that property (Durán and Formanek 2018, p. 15). In this way, robustness aligns with the notion of consilience, as discussed here, and can be applied to AI models in biological sciences to strengthen confidence in their explanatory power and predictive accuracy.

As McFadden (2021) states, the essence of science lies in constructing models that facilitate predictions about the world. If the predictions align cohesively, we establish a coherent depiction of nature. Evidently, coherence does not equate to correctness; rather, it signifies that we have not falsified our model yet. The true nature of reality will always be beyond our comprehension. This elusiveness is why our scientific hypotheses continuously orbit around the core of something fundamentally unknowable. In this sense, AI models represent a modern embodiment of Whewell's (1847) consilience of inductions. They begin with narrow goals and end up explaining broad, diverse phenomena. This pattern of generalization is a signal of epistemic robustness. The consilient behavior of AI models suggests they are capturing something real about the world's structure.

## 6. Conclusion

Schoch et al. (2024) introduced a definition of computational reproducibility that incorporates both the agent and computational environment. The authors suggest solutions such as allowing restricted third-party access or using less expensive equipment, sharing code and datasets, using research compendiums, and providing a declarative description of the computational environment used in the research. Although directly related to reproducibility in computational social science, these conclusions are also worth considering in AI-based biology research.

Sandfort et al. (2024), Santos and Gois (2025), and Pollock et al. (2025) discussed the potential of AI to transform environmental assessment practices, questioning whether biologists are prepared for these changes. Concerns regarding the future integration of AI in biological sciences, such as the quality and integrity of data collected for training AI models, the need to reevaluate educational and training programs to embrace these novel technologies, and responsibilities related to data collection and interpretation should not be neglected (Davinack 2023, Santos and Gois 2023, 2025). All these concerns relate to the central issue of reproducibility, emphasizing the necessity for careful, transparent, and standardized processes.

Ironically, having vast amounts of high-quality data does not guarantee reproducibility. In some cases, it has made it even more difficult. Improving the reproducibility of scientific research, especially in complex fields like ecology, biological systematics, and molecular biology, demands a multifaceted approach grounded in empirical evidence and methodological rigor, standardized protocols, shared databases, and a culture of transparency. Adopting these practices helps manage interpretation, ensuring credible and replicable research findings. However, pursuing reproducibility should not hinder scientific progress, as its success depends on the nature of the research.

In particular, integrating AI into research introduces opportunities and complexities that we must carefully manage. The nature of AI models can challenge the reproducibility of results, as variations in input data, model parameters, or computational environments can lead to divergent outcomes. We highlight the need for standardization and documentation in AI-driven research to ensure the results can be replicated. As biologists increasingly rely on computer code and AI-based tools for data collection and analysis, we emphasize Maitner et al. (2024)'s encouragement of maintaining well-documented code in version-controlled public repositories and public archives.

High-quality datasets are crucial for meaningful scientific insights and hypothesis testing, particularly in AI-driven research. Careful curation enhances their value and fosters scientific progress. Scientific reproducibility remains a crucial guideline in research, but we agree with Hull (1988, p. 26): "Scientists are not objective all the time."

While striving for reproducibility, researchers must embrace divergence due to the subjective nature of scientific work. In this context, the concept of consilience—where independent lines of evidence, sometimes even unrelated at first glance, converge towards a shared conclusion—provides a more flexible and realistic framework for scientific validation. This is especially valid considering AI models, which often show consilience by generating explanatory power across domains far removed from their original training and goals. Integrating consilience with traditional reproducibility ensures scientific progress even when perfect replicability is unattainable. In a world where perfect reproducibility is unattainable, the difficulties in reproducing results should not prevent us from proposing bold hypotheses or unorthodox interpretations.

**Acknowledgments**

This study was partly financed by CNPq, Brazil (304027/2022-7, CMDS), Wild Animal Initiative (Grant Number W-8BEN) and Coordenação de Aperfeiçoamento de Pessoal de Nível Superior, Brazil (CAPES) Finance Code 001 (MPA).